\documentclass[prd,amsmath,amssymb,reprint,preprintnumbers,nofootinbib,superscriptaddress]{revtex4-1}
              
\pdfoutput=1

\usepackage{amsmath}
\usepackage{amsfonts}
\usepackage{amssymb}
\usepackage{mathrsfs}
\usepackage{graphicx}
\usepackage{color}
\usepackage[dvipsnames]{xcolor}
\usepackage{longtable}
\usepackage{bm}
\usepackage{blindtext}
\usepackage{wasysym}
\usepackage{hyperref}
\hypersetup{colorlinks=true,allcolors=blue}
\usepackage[normalem]{ulem}
\usepackage{lineno}
\usepackage{multirow,bigdelim}
\usepackage{adjustbox}
\usepackage{array}
 \usepackage{makecell}

\newcommand\Fontx{\fontsize{10}{12}\selectfont}

\begin{document}
 
\title{Constraints on flavor-dependent long-range interactions
	of high-energy astrophysical neutrinos\\\vskip0.2cm
\Fontx{``Contribution to the 25th International Workshop on Neutrinos from Accelerators''}}
\author{Sanjib Kumar Agarwalla}
\affiliation{Institute of Physics, Sachivalaya Marg, Sainik School Post, Bhubaneswar 751005, India}
\affiliation{Homi Bhabha National Institute, Training School Complex, Anushakti Nagar, Mumbai 400094, India}
\author{Mauricio Bustamante}
\affiliation{Niels Bohr International Academy, Niels Bohr Institute, University of Copenhagen, DK-2100 Copenhagen, Denmark}
\author{Sudipta Das}
\affiliation{Institute of Physics, Sachivalaya Marg, Sainik School Post, Bhubaneswar 751005, India}
\affiliation{Homi Bhabha National Institute, Training School Complex, Anushakti Nagar, Mumbai 400094, India}
\affiliation{Department of Physics and Astronomy,  University of Iowa, Iowa City, IA 52242, USA}
\author{Ashish Narang}
\affiliation{Institute of Physics, Sachivalaya Marg, Sainik School Post, Bhubaneswar 751005, India}
\affiliation{Homi Bhabha National Institute, Training School Complex, Anushakti Nagar, Mumbai 400094, India}

\preprint{NuFact 2024-35}

\date{\today}

\begin{abstract}
 Astrophysical neutrinos
 with energy in the TeV-PeV range traverse megaparsecs (Mpc) to gigaparsecs (Gpc) scale distances before they reach the Earth. Tiny physics effects that get  accumulated over these large propagation paths during their journey may become observable at the detector. If there is some new interaction between neutrinos and the background matter, that can potentially affect the propagation of the astrophysical neutrinos. One such possible case is the flavor-dependent long-range interaction of neutrinos, which can affect the standard neutrino flavor transition, modifying the flavor composition of the astrophysical neutrinos at Earth. Using the present-day and future projection of the flavor-composition measurements of IceCube and IceCube-Gen2 along with the present and future measurement of the oscillation parameters, we explore the sensitivity of these experiments to probe long-range neutrino interaction with matter.

\begin{center} (presented by Sudipta Das)\end{center}
 
\end{abstract}


\maketitle


\section{Introduction} 
High-energy astrophysical neutrinos offer exceptional opportunities to probe fundamental physics. Generated within the intense, high-energy environments of galactic and extra-galactic sources, these neutrinos traverse distances of the order of gigaparsecs before arriving at Earth. Due to their high energy and large propagation distance, minute physics effects can accumulate during their propagation and become detectable by neutrino telescopes at Earth.

In this work, we explore the possible flavor-dependent long-range interaction~(LRI) of neutrinos with the matter in the local and distant Universe~\cite{Joshipura:2003jh,Bandyopadhyay:2006uh,Gonzalez-Garcia:2006vic}. This interaction can occur within the framework of $U(1)'$ extensions of the Standard Model~(SM) gauge group. These models necessarily introduce a new neutral gauge boson $Z'$, which, if ultra-light, mediates the long-range interaction of neutrinos. The presence of LRI in Nature would influence the flavor fraction of astrophysical neutrinos reaching the Earth. We discuss how neutrino flavor composition measurements in IceCube and projected measurements in the IceCube-Gen2 can probe various possible scenarios of LRI. Note that a part of the results discussed in this article are already discussed in ref.~\cite{Agarwalla:2023sng}.
 \vspace{-0.1cm}
\section{Neutrino-matter interactions in $U(1)'$ Models}
\label{sec:U(1)}
\vspace{-0.1cm}
In the SM framework, neutrinos interact with ordinary matter particles {\it via} $W$ boson-mediated charge-current~(CC) interactions and $Z$ boson-mediated neutral-current~(NC) interactions. However, a new neutrino-matter interactions is possible in various beyond the Standard Model~(BSM) proposals. In this work, we discuss one such possibility, the gauged $U(1)'$ models, which may introduce long-range interactions of astrophysical neutrinos.

The SM extension with a new $U(1)'$ gauge group introduces a new neutral gauge boson, $Z'$, that mediates the interaction between neutrinos and ordinary matter fermions~($e,u$, and $d$) naturally. The relevant terms in the interaction Lagrangian are
\begin{multline}
	\label{eq:Lagrangian}
	\mathcal{L}_{Z'}^\text{matter}
	= -g'\big( a_u\, \bar{u}\gamma^\alpha u
	+ a_d\, \bar{d}\gamma^\alpha d
	+ a_e\, \bar{e}\gamma^\alpha e
	\\
	+ b_e \,\bar{\nu}_e\gamma^\alpha P_L \nu_e
	+ b_\mu\,\bar{\nu}_\mu\gamma^\alpha P_L \nu_\mu
	+ b_\tau\, \bar{\nu}_\tau \gamma^\alpha P_L \nu_\tau \big)  Z'_\alpha\,,
\end{multline}
where $a_{e,u,d}$ and $b_{e,\mu,\tau}$ are the $U(1)'$ charges associated with the matter fermions and neutrinos, respectively, whose values depend on the symmetry of the new $U(1)'$ gauge group. However, an arbitrary symmetry of the $U(1)'$ group may lead to anomalies in the model and certain combinations of baryon and lepton number symmeteries are anomaly free. 
\begin{table}[tbp]
	\centering
	\catcode`!=\active\def!{\hphantom{-}}
	\catcode`?=\active\def?{\hphantom{0}}
	\begin{tabular}{@{\rule[-2.1mm]{0pt}{2mm}}|c|cccccc|}
		\hline
		Model
		& $a_u$ & $a_d$ & $a_e$ & $b_e$ & $b_\mu$ & $b_\tau$
		\\
		\hline
		$L - 3L_\mu$
		& 0 & 0 & $!1$ & $!1$ & $-2$ & $!1$
		\\
		$L - 3L_\tau$
		& 0 & 0 & $!1$ & $!1$ & $!1$ & $-2$
		\\
		$L_e - L_\mu$
		& $0$ & $0$ & $!1$ & $!1$ & $-1$ & $!0$
		\\
		$L_e - L_\tau$
		& $0$ & $0$ & $!1$ & $!1$ & $!0$ & $-1$
		\\
		\hline
	\end{tabular}
\caption{List of anomaly-free symmetries for the $U(1)'$ gauge group. The parameters $a_l$ and $b_l$ ($l = e, \mu, \tau$) are the $U(1)^\prime$ charges associated with the each fermion and neutrino, respectively. In these models, $L= L_e+L_\mu+L_\tau$.}
\label{tab:list_of_models}
\end{table}
In this  work, we consider some anomaly-free models involving only lepton numbers. In Table~\ref{tab:list_of_models}, we list them along with the associated $U(1)'$ charges for ordinary fermions and neutrinos. The  new neutrino-matter interaction potential that arises from these models has the form
\begin{equation}
	\label{eq:eps-nsi-fermion}
	V^f_{\alpha\beta}  = \delta_{\alpha\beta}\, a_f\, b_\alpha\, V_0\,,
\end{equation} 
where $V_0$ is the strength of the interaction potential, a function of coupling strength $g'$ and mass of $Z'$, denoted by $m_Z'$.

Note that a sub-leading component of the new interaction potential may arise from the mixing between $Z'$ and the SM gauge boson $Z$~\cite{Agarwalla:2023sng,Singh:2023nek,Agarwalla:2024ylc}. However, in this work, we only consider the interactions at the tree-level mediated solely by $Z'$.


\section{Long-range interactions of Astrophysical neutrinos} 

\begin{figure*}[tbp]
	\centering
	\includegraphics[width=0.45\linewidth]{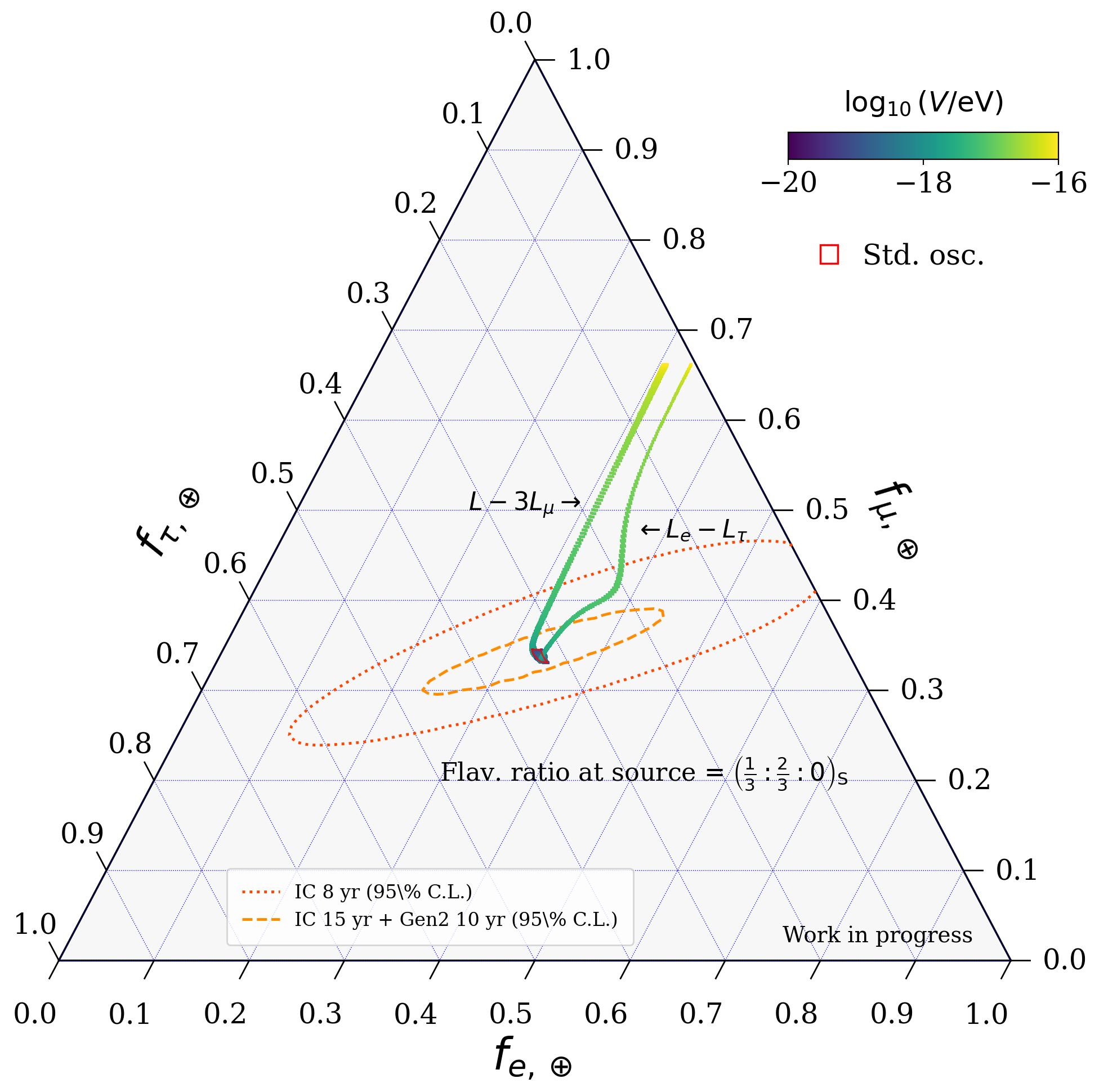}
	\includegraphics[width=0.45\linewidth]{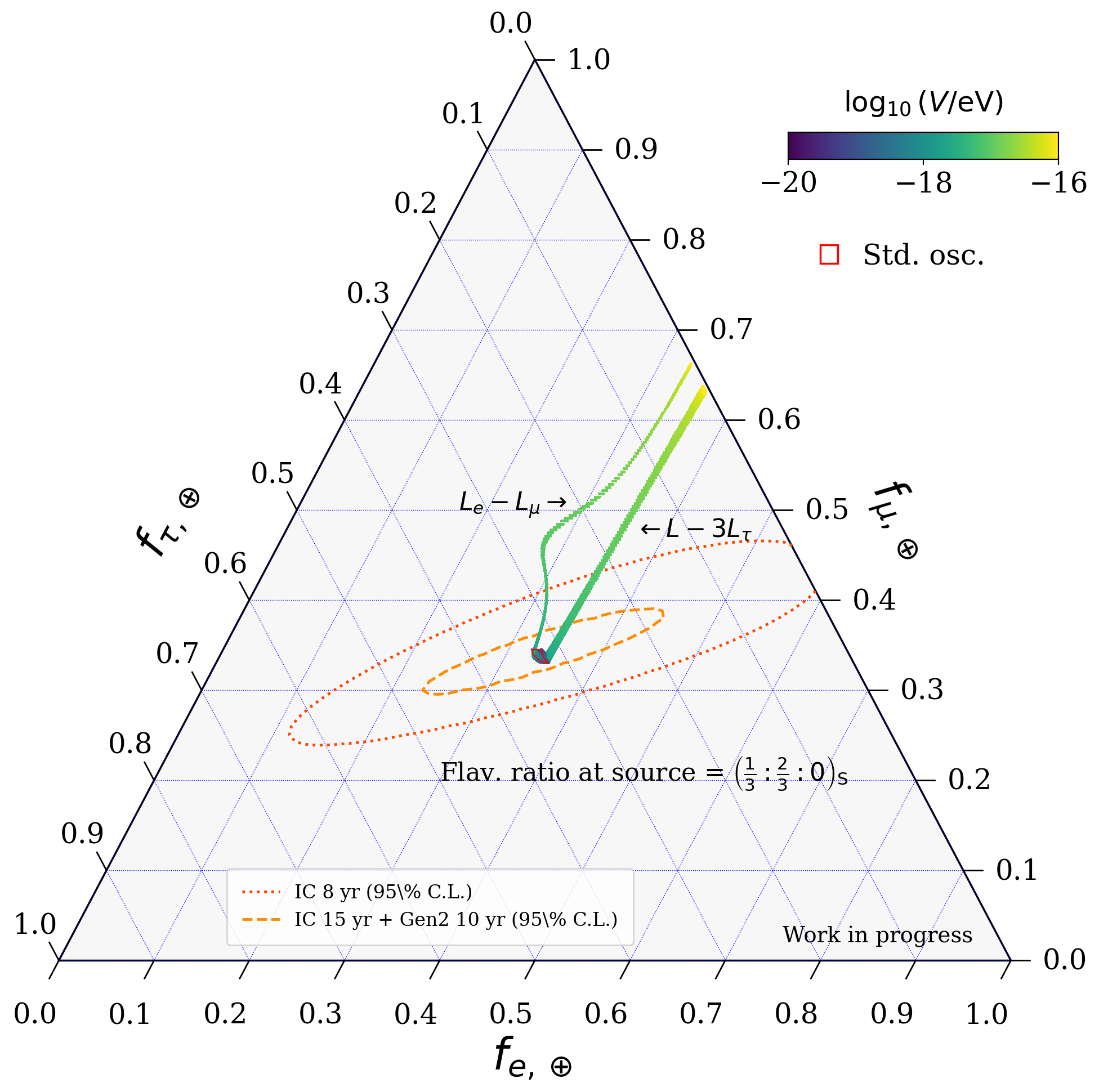}
	\caption{Neutrino flavor composition at Earth as a function of LRI potential induced by all the anomaly-free models considered in this work. In the left~(right) panel, the two  bands correspond to  $L-3L_\mu$ and $L_e-L_\tau$ ($L-3L_\tau$ and $L_e-L_\mu$), as labelled in the plot. To compute the flavor composition at Earth (see eq.~(\ref{eq:flavor_comp})), we use the value of oscillation parameters from refs.~\cite{Esteban:2020cvm,NuFIT}, in their 1$\sigma$ allowed ranges for normal mass ordering~(NMO). We use constant neutrino energy $E=100$~TeV. The two contours correspond to the flavor composition estimated/projected from IceCube 8 years of high-energy starting events~(HESE)\,+\,through-going muons, and IceCube\,+\,IceCube-Gen2 with a total of 25 years of exposure, respectively.}
	\label{fig:flav_comp}
\end{figure*}

 Astrophysical neutrinos undergo flavor oscillations while propagating from the source to the Earth, which modifies the fraction of each flavor in the neutrino flux that reaches Earth, {\it i.e.}, the flavor composition. In general, these neutrinos propagate in vacuum. In such a scenario, the neutrino flavor transition probability is given by $P_{\alpha\beta}=\sum_{i=1,2,3} |U_{\alpha i}|^2 |U_{\beta i}|^2$, where $U_{\alpha i}$ and $U_{\beta i}$ are the elements of the standard three-neutrino PMNS mixing matrix. Note that, in this expression, the effect of the mass-squared difference is averaged out as $L/E>>1$, where $L$ is the propagation distance and $E$ is the neutrino energy. Present measurements of the neutrino mixing parameters allow us to estimate the flavor composition at Earth, $f_{\oplus} = (f_{e,\oplus},f_{\mu,\oplus},f_{\tau,\oplus})$, for a known flavor composition at source $f_{S} = (f_{e,S},f_{\mu,S},f_{\tau,S})$,
 \begin{equation}
\label{eq:flavor_comp}
f_{\beta,\oplus} = \sum_{\alpha = e,\mu,\tau} P_{\beta\alpha} f_{\alpha,S}\,.
\end{equation}
This work considers the pion decay scenario as the neutrino production mechanism at the source, where pions are produced from the collision of high-energy protons with background protons or photons. Produced pions decay to produce one $\nu_\mu$ or $\bar{\nu}_\mu$ and $\mu^{+}$ or $ \mu^{-}$, which further decay to produce one $\bar{\nu}_\mu$ or $\nu_\mu$ and one $\nu_e$ or $\bar{\nu}_e$, along with an $e^+$ or $e^-$. Overall, the flavor composition at the source becomes $(1/3:2/3:0)$. In this case, flavor composition at the Earth calculated using eq.~(\ref{fig:flav_comp}) is around $f_\oplus\approx (1:1:1)$. 

As mentioned earlier, the presence of a possible neutrino-matter LRI with sufficient strength may influence the neutrino flavor composition at Earth, and such interaction may arise from gauged $U(1)'$ models in the ultralight mass limit of $Z'$. The interaction range, in this case, is inversely proportional to $m_Z'$. As a result, the interaction can be sourced by the matter particles inside distant astrophysical bodies that fall within the interaction range. For a neutrino propagating through the universe, 
the strength of the potential due to the $Z'$-mediated interaction~(as defined in eq.~(\ref{eq:eps-nsi-fermion})) is given by
\begin{equation}
\label{eq:V0}
V_0=g'^2\frac{e^{-m_Z' s}}{4\pi s}\times N_f\,,
\end{equation}

where $s$ denotes the distance between the neutrino and the source particle, and $N_f$ is the number density of the fermion $f$. For a neutrino at the surface of the Earth, the interaction potential can be sourced by the fermions from the Earth~($\oplus$), the Moon~($\leftmoon$), the Sun~($\astrosun$), the Milkiway~(MW), and cosmological fermions, depending on the value of $m_Z'$. The smaller the mass of $m_Z'$ would lead to longer interaction range. The total interaction potential is the sum of all contributions from all the celestial bodies, {\it i.e.} 
$V_{\rm LRI} = V^\oplus + V^{\leftmoon} + V^{\astrosun} + V^{\rm MW} + \langle V^{\rm cos} \rangle$.
where, in the last term, we average over the redshift $z$.
To calculate potential from each of the celestial bodies, we follow the formalism as discussed in refs.~\cite{Bustamante:2018mzu,Agarwalla:2023sng}.

In the  presence of LRI, the neutrino propagation Hamiltonian can be written as
\begin{equation}
\label{equ:hamiltonian_tot}
\mathbf{H}
=
\mathbf{H}_{\rm vac}
+
\mathbf{V}_{\rm mat}
+
\mathbf{V}_{\rm LRI} \;.
\end{equation}
 The first term, $\mathbf{H}_{\rm vac}= 
U\cdot
{\rm diag}(0, \frac{\Delta m^2_{21}}{2E}, \frac{\Delta m^2_{31}}{2E})
\cdot U^{\dagger}$, is the usual neutrino Hamiltonian in vacuum, where $U$ denotes thr neutrino mixing matrix, and $\Delta m^2_{ij}$ are neutrino mass-squared differences.
The contribution from the standard neutrino-matter interaction comes from
$
\mathbf{V}_{\rm mat}
=
{\rm diag}(V_{\rm CC}, 0, 0)$, where $V_{\rm CC}$ is the standard matter potential due to the CC interaction.
The last term  $\mathbf{V}_{\rm LRI}$ in eq.~(\ref{equ:hamiltonian_tot}) brings in the effect of the new LRI, whose elements are defined in eq.~(\ref{eq:eps-nsi-fermion}), suggesting that the matrix is diagonal.
The structure of the $\mathbf{V}_{\rm LRI}$ sourced by the electrons for the four $U(1)'$ models that we consider in this work are:
\begin{align}
&\mathbf{V}^{(L-3L_\mu)}_{\rm LRI} = \begin{pmatrix}
0 & 0 & 0\\
0 & 3V & 0\\
0 & 0 & 0
\end{pmatrix}\,,
\hspace{0.5cm}
\mathbf{V}^{(L-3L_\tau)}_{\rm LRI} = \begin{pmatrix}
0 & 0 & 0\\
0 & 0 & 0\\
0 & 0 & 3V
\end{pmatrix}\,,\\\nonumber
&\mathbf{V}^{(L_e-L_\mu)}_{\rm LRI} =
\begin{pmatrix}
V & 0 & 0\\
0 & -V & 0\\
0 & 0 & 0
\end{pmatrix}\,,
\mathbf{V}^{(L_e-L_\tau)}_{\rm LRI} =
\begin{pmatrix}
V & 0 & 0\\
0 & 0 & 0\\
0 & 0 & -V
\end{pmatrix}\,.
\end{align}

 We calculate the flavor transition probabilities by diagonalizing the Hamiltonian defined in eq.~(\ref{equ:hamiltonian_tot}). We use the expression  
 $
 \bar{P}_{\alpha\beta}
 =
 \sum^3_{i=1}|U^m_{\alpha i}|^2|U^m_{\beta i}|^2 
 $ for the oscillation probabilities in the presence of LRI,
 where $U^m$ is the new mixing matrix that diagonalizes $\mathbf{H}_{\rm tot}$.
 Now, we estimate the flavor composition at Earth using eq.(~\ref{eq:flavor_comp}) using the pion decay scenario at the source. For our calculation, we use the value of the oscillation parameters from refs.~\cite{Esteban:2020cvm,NuFIT}, in their $1\sigma$ allowed ranges.

In figure~\ref{fig:flav_comp}, we show the neutrino flavor composition at Earth for a constant neutrino energy $E=100$ TeV in the presence of LRI for the four models listed in table~\ref{tab:list_of_models}.  The strength of the interaction potential is varied in the range [$10^{-22},10^{-16}$]~eV, relevant for our analysis. We observe that in the limit ~($V_{\alpha\beta} << \Delta m^2_{21}/2E$), the effect of LRI is negligible and the expected $f_\oplus$ stays near standard expectation~{\it i.e.,} (1:1:1). However, as the strength of $V_{\alpha\beta}$ starts to increase, for all the models, the flavor composition at Earth starts to deviate from the standard expectation. In the limit~($V_{\alpha\beta} >> \Delta m^2_{21}/2E$), the contribution from the LRI potential dominates, oscillations get suppressed, and the flavor composition at Earth tends to remain in its initial ratio, {\it i.e.} $(1/3:2/3:0)$. Note that for $L-3L_\mu$ case, the flavor ratio at Earth does not exactly saturate to the initial ratio. It happens because the flavor transition probabilities do not attain the extreme values, {\it i.e.} 0 or 1, even for the large value of LRI potential.

\vspace{-0.2cm}
\section{Simulation details}

We contrast the calculated flavor compositions at Earth with estimated/projected flavor compositions as measured by various neutrino telescopes (shown by the contours in figure~\ref{fig:flav_comp}). First, we consider IceCube flavor composition estimates with 8 years of HESE and through-going muons~\cite{Song:2020nfh}. For our future projections, we consider 15 years of IceCube HESE\,+\,through-going muon data and 10 years of IceCube-Gen2~\cite{IceCube-Gen2:2020qha}. Since this detector does not distinguish between neutrinos and antineutrinos, we average the flavor composition calculated using eq.~(\ref{eq:flavor_comp}), over neutrino and antineutrino modes, {\it i.e.} $f_\oplus = \frac{f_\oplus(\nu)+f_\oplus(\bar{\nu})}{2}$.
 Also, we average the computed flavor composition over neutrino energies in the range of 25 TeV to 2.8 PeV, where IceCube flavor measurements satisfy an unbroken power law~\cite{IceCube:2015gsk}.
For the oscillation parameters, we consider the present 1$\sigma$ allowed ranges on the neutrino mass-mixing parameters from ref.~\cite{Esteban:2020cvm,NuFIT}, as well as the future improvements from DUNE, T2HK, and JUNO.  

We compute the posterior probability distribution of the LRI potential for each model, which is defined as 
\begin{equation}
\label{equ:posterior}
\mathcal{P}\left(V_{\alpha \beta}\right)
=
\int d\pmb{\vartheta}
\mathcal{L}
\left(\langle\pmb{f}_\oplus\left(V_{\alpha \beta}, \pmb{\vartheta}\right)\rangle\right)
\pi(\pmb{\vartheta})
\pi\left(V_{\alpha \beta}\right) \;,
\end{equation}
 where $\mathcal{L}
 \left(\langle\pmb{f}_\oplus\left(V_{\alpha \beta}, \pmb{\vartheta}\right)\rangle\right)
 $ is the likelihood of having the measured flavor composition $\langle \pmb{f}_{\oplus} \rangle$. $\pi(\pmb{\vartheta})$ is the priors on the oscillation parameters $\pmb{\vartheta} \equiv (\theta_{12}, \theta_{23}, \theta_{13}, \delta_{\rm CP})$, for which we use Gaussian priors which are centered at their best-fit values as given in ref.~\cite{Esteban:2020cvm,NuFIT}, assuming normal mass ordering~(NMO) of neutrino. For the prior on $V_{\alpha\beta}$, $\pi(V_{\alpha\beta})$, we use uniform priors in the range [$10^{-24},10^{-16}$]~eV. 
 After computing the posterior probability distribution, we compute the upper limit of the long-range potential for each $U(1)'$ model at 95\% confidence level~(C.L.). 
 
\vspace{-0.2cm}
\section{Results}
\vspace{-0.2cm}
\begin{figure*}[tbp]
	\centering
	\includegraphics[width= 0.44\textwidth]{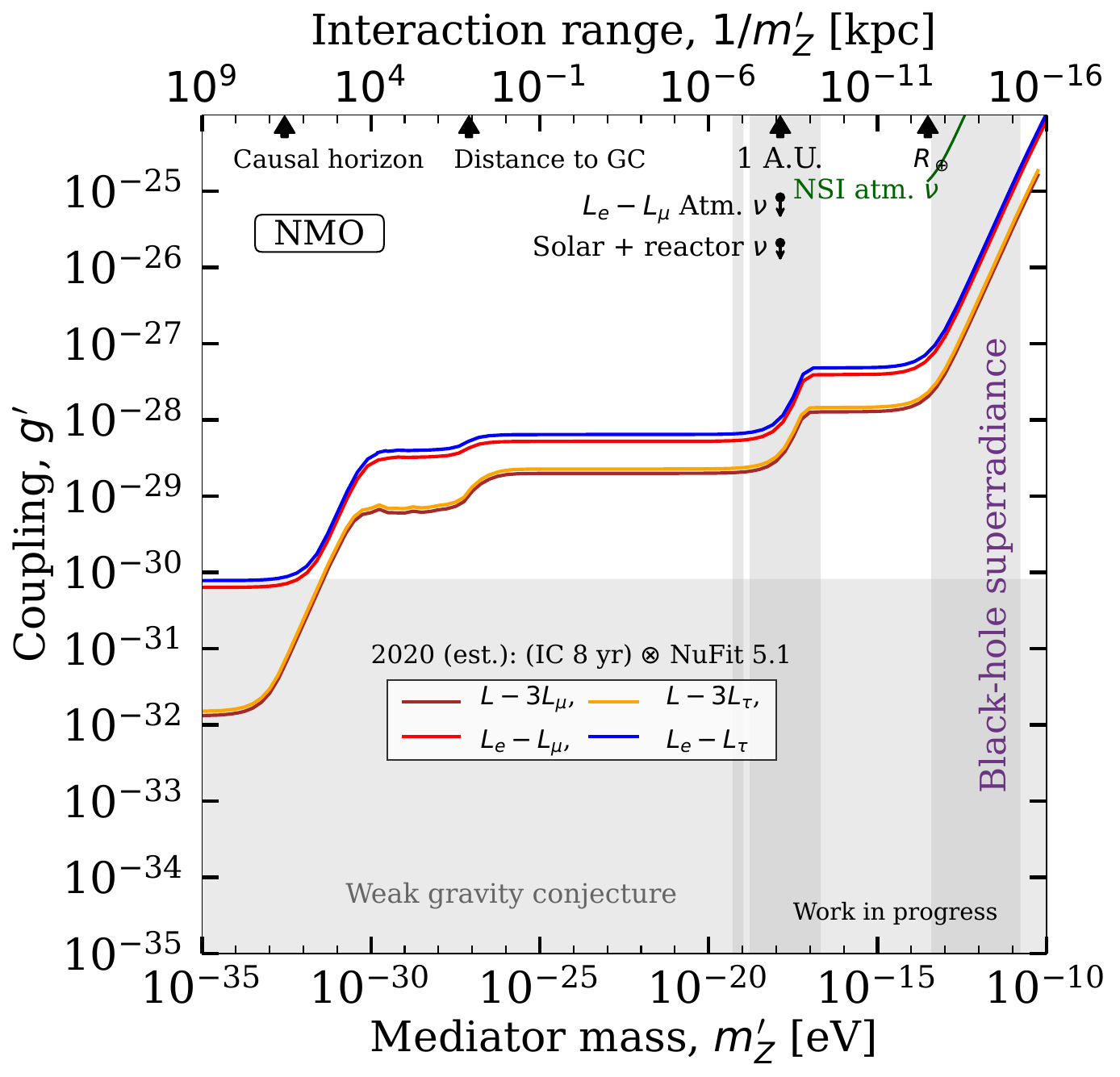}
	\includegraphics[width= 0.44\textwidth]{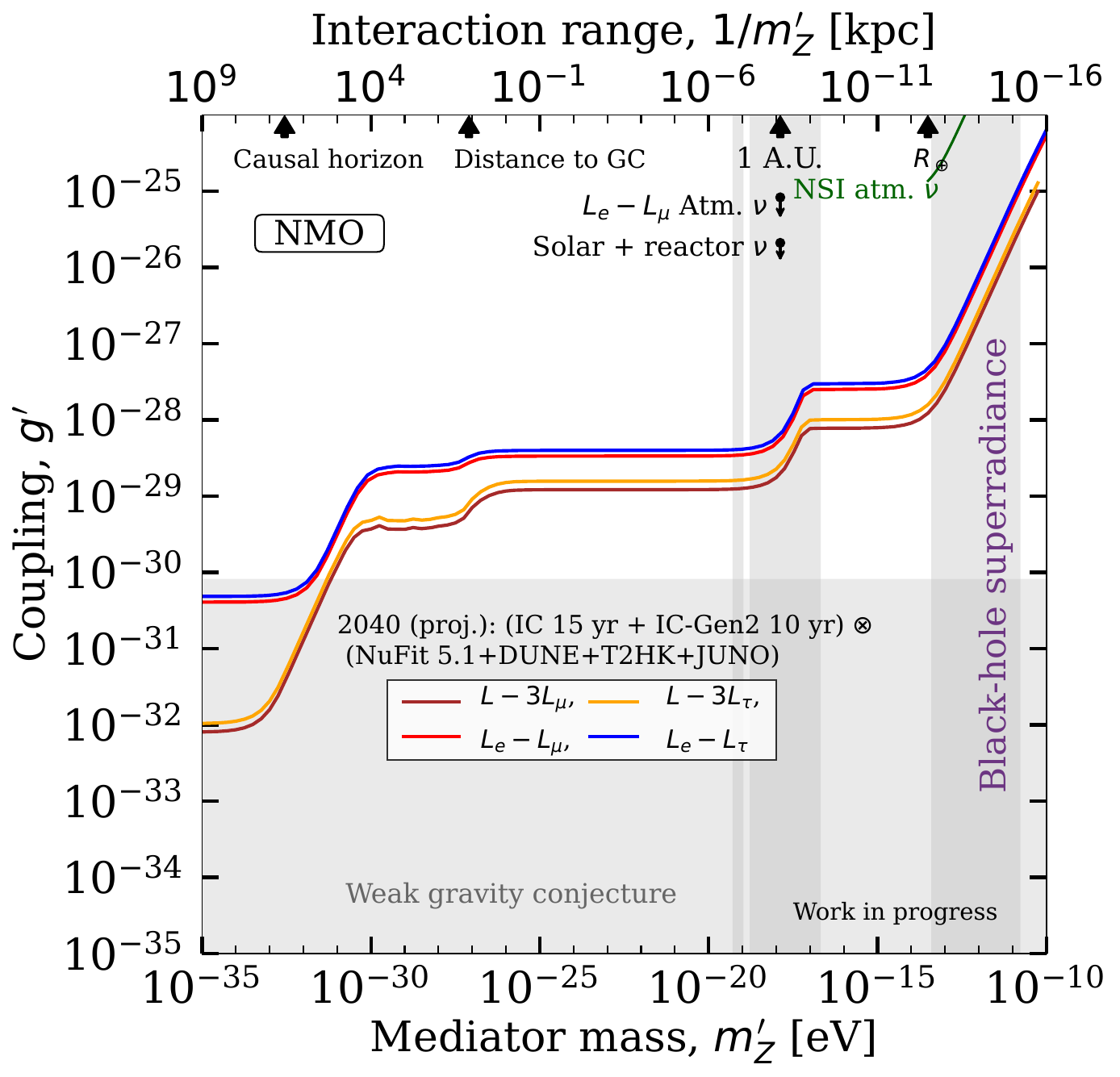}
	\caption{95\% C.L. limits on the coupling strength of the LRI arising from gauged $U(1)'$ models with various anomaly-free symmetries as a function of mass of $Z'$. The left panel shows estimated limits from IceCube with the 8 years of HESE+through going muon sample, while the right panel corresponds to the projected 10 years of IceCube-Gen2 event sample along with 15 years of IceCube projected data. For comparison, we also show the existing limits on long-range interactions from a global fit of oscillation data~\cite{Coloma:2020gfv}, atmospheric neutrinos~\cite{Joshipura:2003jh}, solar and reactor neutrinos~\cite{Bandyopadhyay:2006uh}, and non-standard interactions (NSI)~\cite{Super-Kamiokande:2011dam, Ohlsson:2012kf, Gonzalez-Garcia:2013usa}. Also, we show the indirect limits from  
		black-hole superradiance (90\% C.L.)~\cite{Baryakhtar:2017ngi} and the weak gravity conjecture~\cite{Arkani-Hamed:2006emk}, assuming a lightest neutrino mass of $0.01$~eV. }
	\label{fig:limits_g_m}
\end{figure*}

\begin{table}[tbp]
	\centering
	\begin{tabular}{ | c | *{4}{>{\centering\arraybackslash}p{3.0cm} |}}
		\hline
		\multirow{3}{*}{Models} &
		\multicolumn{2}{c|}{Upper limit (95\%~C.L.) on potential [$10^{-19}$~eV]} \\
		\cline{2-3}
		&
		\multicolumn{1}{c|}{IC 8 yr} &
		\multicolumn{1}{c|}{IC 15 yr + Gen2 10 yr}  \\
		\hline
		$L - 3L_\mu$ 
		&  3.20 &  1.19  \\
		$L-3L_\tau$ & 3.11 & 1.11 \\
		$L_e-L_\mu$ & 4.41 & 1.69 \\
		$L_e-L_\tau$ & 1.79 & 0.731 \\
		\hline
	\end{tabular}
	\caption{95 \% C.L. upper limits on the LRI potential arising from the anomaly-free $U(1)'$ models considered in this work.}
	\label{tab:bounds}
\end{table}
In table~\ref{tab:bounds}, we list the upper limit on the LRI potential at 95\% C.L. for each model. We observe that the addition of the projected measurements from IceCube-Gen2 with 8 years of IceCube estimates leads to a significant improvement in the results.

Now, in figure~\ref{fig:limits_g_m}, we plot the limits mentioned in table~\ref{tab:bounds} in the ($g'$-$m_Z'$) plane. We use eq.(~\ref{eq:V0}) to plot the contours for various $U(1)'$ symmetries as shown by colored lines.
The step-like behavior in each of the models at particular values of the $m'_{Z}$ is due to the inclusion of the contribution from electrons inside the celestial bodies like the Sun, the galactic center where matter density is high, and when the interaction length reaches the causal horizon.

\vspace{-0.2cm}
\section{Conclusions}
\vspace{-0.2cm}

This work explores the possible impact of flavor-dependent neutrino-matter long-range interactions (LRI) on the flavor composition of astrophysical neutrinos at Earth. We test $U(1)'$ models with a selection of anomaly-free symmetries, which may induce these LRI. We use flavor-composition estimates from IceCube and also future projections from IceCube-Gen2 to place limits on LRI. IceCube estimates put stringent constraints on the $U(1)'$ models considered in this work, improving over the existing limits. Our IceCube-Gen2 projections further improve these results, approximately by a factor of two.

\begin{acknowledgments}
We acknowledge financial support from the DAE, DST, DST-SERB, Govt. of India, INSA, and the USIEF. Numerical simulations 
are performed using the “SAMKHYA: High-Performance Computing Facility” at the Institute of Physics, Bhubaneswar, India. 
\end{acknowledgments}

\bibliographystyle{apsrev4-1}
\bibliography{refer-lri.bib}

\begin{thebibliography}{18}%
\makeatletter
\providecommand \@ifxundefined [1]{%
 \@ifx{#1\undefined}
}%
\providecommand \@ifnum [1]{%
 \ifnum #1\expandafter \@firstoftwo
 \else \expandafter \@secondoftwo
 \fi
}%
\providecommand \@ifx [1]{%
 \ifx #1\expandafter \@firstoftwo
 \else \expandafter \@secondoftwo
 \fi
}%
\providecommand \natexlab [1]{#1}%
\providecommand \enquote  [1]{``#1''}%
\providecommand \bibnamefont  [1]{#1}%
\providecommand \bibfnamefont [1]{#1}%
\providecommand \citenamefont [1]{#1}%
\providecommand \href@noop [0]{\@secondoftwo}%
\providecommand \href [0]{\begingroup \@sanitize@url \@href}%
\providecommand \@href[1]{\@@startlink{#1}\@@href}%
\providecommand \@@href[1]{\endgroup#1\@@endlink}%
\providecommand \@sanitize@url [0]{\catcode `\\12\catcode `\$12\catcode
  `\&12\catcode `\#12\catcode `\^12\catcode `\_12\catcode `\%12\relax}%
\providecommand \@@startlink[1]{}%
\providecommand \@@endlink[0]{}%
\providecommand \url  [0]{\begingroup\@sanitize@url \@url }%
\providecommand \@url [1]{\endgroup\@href {#1}{\urlprefix }}%
\providecommand \urlprefix  [0]{URL }%
\providecommand \Eprint [0]{\href }%
\providecommand \doibase [0]{http://dx.doi.org/}%
\providecommand \selectlanguage [0]{\@gobble}%
\providecommand \bibinfo  [0]{\@secondoftwo}%
\providecommand \bibfield  [0]{\@secondoftwo}%
\providecommand \translation [1]{[#1]}%
\providecommand \BibitemOpen [0]{}%
\providecommand \bibitemStop [0]{}%
\providecommand \bibitemNoStop [0]{.\EOS\space}%
\providecommand \EOS [0]{\spacefactor3000\relax}%
\providecommand \BibitemShut  [1]{\csname bibitem#1\endcsname}%
\let\auto@bib@innerbib\@empty
\bibitem [{\citenamefont {Joshipura}\ and\ \citenamefont
  {Mohanty}(2004)}]{Joshipura:2003jh}%
  \BibitemOpen
  \bibfield  {author} {\bibinfo {author} {\bibfnamefont {A.~S.}\ \bibnamefont
  {Joshipura}}\ and\ \bibinfo {author} {\bibfnamefont {S.}~\bibnamefont
  {Mohanty}},\ }\href {\doibase 10.1016/j.physletb.2004.01.057} {\bibfield
  {journal} {\bibinfo  {journal} {Phys. Lett. B}\ }\textbf {\bibinfo {volume}
  {584}},\ \bibinfo {pages} {103} (\bibinfo {year} {2004})},\ \Eprint
  {http://arxiv.org/abs/hep-ph/0310210} {arXiv:hep-ph/0310210} \BibitemShut
  {NoStop}%
\bibitem [{\citenamefont {Bandyopadhyay}\ \emph {et~al.}(2007)\citenamefont
  {Bandyopadhyay}, \citenamefont {Dighe},\ and\ \citenamefont
  {Joshipura}}]{Bandyopadhyay:2006uh}%
  \BibitemOpen
  \bibfield  {author} {\bibinfo {author} {\bibfnamefont {A.}~\bibnamefont
  {Bandyopadhyay}}, \bibinfo {author} {\bibfnamefont {A.}~\bibnamefont
  {Dighe}}, \ and\ \bibinfo {author} {\bibfnamefont {A.~S.}\ \bibnamefont
  {Joshipura}},\ }\href {\doibase 10.1103/PhysRevD.75.093005} {\bibfield
  {journal} {\bibinfo  {journal} {Phys. Rev. D}\ }\textbf {\bibinfo {volume}
  {75}},\ \bibinfo {pages} {093005} (\bibinfo {year} {2007})},\ \Eprint
  {http://arxiv.org/abs/hep-ph/0610263} {arXiv:hep-ph/0610263} \BibitemShut
  {NoStop}%
\bibitem [{\citenamefont {Gonzalez-Garcia}\ \emph {et~al.}(2007)\citenamefont
  {Gonzalez-Garcia}, \citenamefont {de~Holanda}, \citenamefont {Masso},\ and\
  \citenamefont {Zukanovich~Funchal}}]{Gonzalez-Garcia:2006vic}%
  \BibitemOpen
  \bibfield  {author} {\bibinfo {author} {\bibfnamefont {M.~C.}\ \bibnamefont
  {Gonzalez-Garcia}}, \bibinfo {author} {\bibfnamefont {P.~C.}\ \bibnamefont
  {de~Holanda}}, \bibinfo {author} {\bibfnamefont {E.}~\bibnamefont {Masso}}, \
  and\ \bibinfo {author} {\bibfnamefont {R.}~\bibnamefont
  {Zukanovich~Funchal}},\ }\href {\doibase 10.1088/1475-7516/2007/01/005}
  {\bibfield  {journal} {\bibinfo  {journal} {JCAP}\ }\textbf {\bibinfo
  {volume} {01}},\ \bibinfo {pages} {005} (\bibinfo {year} {2007})},\ \Eprint
  {http://arxiv.org/abs/hep-ph/0609094} {arXiv:hep-ph/0609094} \BibitemShut
  {NoStop}%
\bibitem [{\citenamefont {Agarwalla}\ \emph {et~al.}(2023)\citenamefont
  {Agarwalla}, \citenamefont {Bustamante}, \citenamefont {Das},\ and\
  \citenamefont {Narang}}]{Agarwalla:2023sng}%
  \BibitemOpen
  \bibfield  {author} {\bibinfo {author} {\bibfnamefont {S.~K.}\ \bibnamefont
  {Agarwalla}}, \bibinfo {author} {\bibfnamefont {M.}~\bibnamefont
  {Bustamante}}, \bibinfo {author} {\bibfnamefont {S.}~\bibnamefont {Das}}, \
  and\ \bibinfo {author} {\bibfnamefont {A.}~\bibnamefont {Narang}},\ }\href
  {\doibase 10.1007/JHEP08(2023)113} {\bibfield  {journal} {\bibinfo  {journal}
  {JHEP}\ }\textbf {\bibinfo {volume} {08}},\ \bibinfo {pages} {113} (\bibinfo
  {year} {2023})},\ \Eprint {http://arxiv.org/abs/2305.03675} {arXiv:2305.03675
  [hep-ph]} \BibitemShut {NoStop}%
\bibitem [{\citenamefont {Singh}\ \emph {et~al.}(2023)\citenamefont {Singh},
  \citenamefont {Bustamante},\ and\ \citenamefont {Agarwalla}}]{Singh:2023nek}%
  \BibitemOpen
  \bibfield  {author} {\bibinfo {author} {\bibfnamefont {M.}~\bibnamefont
  {Singh}}, \bibinfo {author} {\bibfnamefont {M.}~\bibnamefont {Bustamante}}, \
  and\ \bibinfo {author} {\bibfnamefont {S.~K.}\ \bibnamefont {Agarwalla}},\
  }\href {\doibase 10.1007/JHEP08(2023)101} {\bibfield  {journal} {\bibinfo
  {journal} {JHEP}\ }\textbf {\bibinfo {volume} {08}},\ \bibinfo {pages} {101}
  (\bibinfo {year} {2023})},\ \Eprint {http://arxiv.org/abs/2305.05184}
  {arXiv:2305.05184 [hep-ph]} \BibitemShut {NoStop}%
\bibitem [{\citenamefont {Agarwalla}\ \emph {et~al.}(2024)\citenamefont
  {Agarwalla}, \citenamefont {Bustamante}, \citenamefont {Singh},\ and\
  \citenamefont {Swain}}]{Agarwalla:2024ylc}%
  \BibitemOpen
  \bibfield  {author} {\bibinfo {author} {\bibfnamefont {S.~K.}\ \bibnamefont
  {Agarwalla}}, \bibinfo {author} {\bibfnamefont {M.}~\bibnamefont
  {Bustamante}}, \bibinfo {author} {\bibfnamefont {M.}~\bibnamefont {Singh}}, \
  and\ \bibinfo {author} {\bibfnamefont {P.}~\bibnamefont {Swain}},\ }\href
  {\doibase 10.1007/JHEP09(2024)055} {\bibfield  {journal} {\bibinfo  {journal}
  {JHEP}\ }\textbf {\bibinfo {volume} {09}},\ \bibinfo {pages} {055} (\bibinfo
  {year} {2024})},\ \Eprint {http://arxiv.org/abs/2404.02775} {arXiv:2404.02775
  [hep-ph]} \BibitemShut {NoStop}%
\bibitem [{\citenamefont {Bustamante}\ and\ \citenamefont
  {Agarwalla}(2019)}]{Bustamante:2018mzu}%
  \BibitemOpen
  \bibfield  {author} {\bibinfo {author} {\bibfnamefont {M.}~\bibnamefont
  {Bustamante}}\ and\ \bibinfo {author} {\bibfnamefont {S.~K.}\ \bibnamefont
  {Agarwalla}},\ }\href {\doibase 10.1103/PhysRevLett.122.061103} {\bibfield
  {journal} {\bibinfo  {journal} {Phys. Rev. Lett.}\ }\textbf {\bibinfo
  {volume} {122}},\ \bibinfo {pages} {061103} (\bibinfo {year} {2019})},\
  \Eprint {http://arxiv.org/abs/1808.02042} {arXiv:1808.02042 [astro-ph.HE]}
  \BibitemShut {NoStop}%
\bibitem [{\citenamefont {Esteban}\ \emph {et~al.}(2020)\citenamefont
  {Esteban}, \citenamefont {Gonz\'alez-Garc\'ia}, \citenamefont {Maltoni},
  \citenamefont {Schwetz},\ and\ \citenamefont {Zhou}}]{Esteban:2020cvm}%
  \BibitemOpen
  \bibfield  {author} {\bibinfo {author} {\bibfnamefont {I.}~\bibnamefont
  {Esteban}}, \bibinfo {author} {\bibfnamefont {M.~C.}\ \bibnamefont
  {Gonz\'alez-Garc\'ia}}, \bibinfo {author} {\bibfnamefont {M.}~\bibnamefont
  {Maltoni}}, \bibinfo {author} {\bibfnamefont {T.}~\bibnamefont {Schwetz}}, \
  and\ \bibinfo {author} {\bibfnamefont {A.}~\bibnamefont {Zhou}},\ }\href
  {\doibase 10.1007/JHEP09(2020)178} {\bibfield  {journal} {\bibinfo  {journal}
  {JHEP}\ }\textbf {\bibinfo {volume} {09}},\ \bibinfo {pages} {178} (\bibinfo
  {year} {2020})},\ \Eprint {http://arxiv.org/abs/2007.14792} {arXiv:2007.14792
  [hep-ph]} \BibitemShut {NoStop}%
\bibitem [{NuF()}]{NuFIT}%
  \BibitemOpen
  \href {http://www.nu-fit.org/} {}\bibinfo {note} {NuFIT 5.1 (2021),
  http://www.nu-fit.org/}\BibitemShut {NoStop}%
\bibitem [{\citenamefont {Song}\ \emph {et~al.}(2021)\citenamefont {Song},
  \citenamefont {Li}, \citenamefont {Arg\"uelles}, \citenamefont {Bustamante},\
  and\ \citenamefont {Vincent}}]{Song:2020nfh}%
  \BibitemOpen
  \bibfield  {author} {\bibinfo {author} {\bibfnamefont {N.}~\bibnamefont
  {Song}}, \bibinfo {author} {\bibfnamefont {S.~W.}\ \bibnamefont {Li}},
  \bibinfo {author} {\bibfnamefont {C.~A.}\ \bibnamefont {Arg\"uelles}},
  \bibinfo {author} {\bibfnamefont {M.}~\bibnamefont {Bustamante}}, \ and\
  \bibinfo {author} {\bibfnamefont {A.~C.}\ \bibnamefont {Vincent}},\ }\href
  {\doibase 10.1088/1475-7516/2021/04/054} {\bibfield  {journal} {\bibinfo
  {journal} {JCAP}\ }\textbf {\bibinfo {volume} {04}},\ \bibinfo {pages} {054}
  (\bibinfo {year} {2021})},\ \Eprint {http://arxiv.org/abs/2012.12893}
  {arXiv:2012.12893 [hep-ph]} \BibitemShut {NoStop}%
\bibitem [{\citenamefont {Aartsen}\ \emph {et~al.}(2021)\citenamefont {Aartsen}
  \emph {et~al.}}]{IceCube-Gen2:2020qha}%
  \BibitemOpen
  \bibfield  {author} {\bibinfo {author} {\bibfnamefont {M.~G.}\ \bibnamefont
  {Aartsen}} \emph {et~al.} (\bibinfo {collaboration} {IceCube-Gen2}),\ }\href
  {\doibase 10.1088/1361-6471/abbd48} {\bibfield  {journal} {\bibinfo
  {journal} {J. Phys. G}\ }\textbf {\bibinfo {volume} {48}},\ \bibinfo {pages}
  {060501} (\bibinfo {year} {2021})},\ \Eprint
  {http://arxiv.org/abs/2008.04323} {arXiv:2008.04323 [astro-ph.HE]}
  \BibitemShut {NoStop}%
\bibitem [{\citenamefont {Aartsen}\ \emph {et~al.}(2015)\citenamefont {Aartsen}
  \emph {et~al.}}]{IceCube:2015gsk}%
  \BibitemOpen
  \bibfield  {author} {\bibinfo {author} {\bibfnamefont {M.~G.}\ \bibnamefont
  {Aartsen}} \emph {et~al.} (\bibinfo {collaboration} {IceCube}),\ }\href
  {\doibase 10.1088/0004-637X/809/1/98} {\bibfield  {journal} {\bibinfo
  {journal} {Astrophys. J.}\ }\textbf {\bibinfo {volume} {809}},\ \bibinfo
  {pages} {98} (\bibinfo {year} {2015})},\ \Eprint
  {http://arxiv.org/abs/1507.03991} {arXiv:1507.03991 [astro-ph.HE]}
  \BibitemShut {NoStop}%
\bibitem [{\citenamefont {Coloma}\ \emph {et~al.}(2021)\citenamefont {Coloma},
  \citenamefont {Gonz\'alez-Garc\'ia},\ and\ \citenamefont
  {Maltoni}}]{Coloma:2020gfv}%
  \BibitemOpen
  \bibfield  {author} {\bibinfo {author} {\bibfnamefont {P.}~\bibnamefont
  {Coloma}}, \bibinfo {author} {\bibfnamefont {M.~C.}\ \bibnamefont
  {Gonz\'alez-Garc\'ia}}, \ and\ \bibinfo {author} {\bibfnamefont
  {M.}~\bibnamefont {Maltoni}},\ }\href {\doibase 10.1007/JHEP01(2021)114}
  {\bibfield  {journal} {\bibinfo  {journal} {JHEP}\ }\textbf {\bibinfo
  {volume} {01}},\ \bibinfo {pages} {114} (\bibinfo {year} {2021})},\ \Eprint
  {http://arxiv.org/abs/2009.14220} {arXiv:2009.14220 [hep-ph]} \BibitemShut
  {NoStop}%
\bibitem [{\citenamefont {Mitsuka}\ \emph {et~al.}(2011)\citenamefont {Mitsuka}
  \emph {et~al.}}]{Super-Kamiokande:2011dam}%
  \BibitemOpen
  \bibfield  {author} {\bibinfo {author} {\bibfnamefont {G.}~\bibnamefont
  {Mitsuka}} \emph {et~al.} (\bibinfo {collaboration} {Super-Kamiokande}),\
  }\href {\doibase 10.1103/PhysRevD.84.113008} {\bibfield  {journal} {\bibinfo
  {journal} {Phys. Rev. D}\ }\textbf {\bibinfo {volume} {84}},\ \bibinfo
  {pages} {113008} (\bibinfo {year} {2011})},\ \Eprint
  {http://arxiv.org/abs/1109.1889} {arXiv:1109.1889 [hep-ex]} \BibitemShut
  {NoStop}%
\bibitem [{\citenamefont {Ohlsson}(2013)}]{Ohlsson:2012kf}%
  \BibitemOpen
  \bibfield  {author} {\bibinfo {author} {\bibfnamefont {T.}~\bibnamefont
  {Ohlsson}},\ }\href {\doibase 10.1088/0034-4885/76/4/044201} {\bibfield
  {journal} {\bibinfo  {journal} {Rept. Prog. Phys.}\ }\textbf {\bibinfo
  {volume} {76}},\ \bibinfo {pages} {044201} (\bibinfo {year} {2013})},\
  \Eprint {http://arxiv.org/abs/1209.2710} {arXiv:1209.2710 [hep-ph]}
  \BibitemShut {NoStop}%
\bibitem [{\citenamefont {Gonz\'alez-Garc\'ia}\ and\ \citenamefont
  {Maltoni}(2013)}]{Gonzalez-Garcia:2013usa}%
  \BibitemOpen
  \bibfield  {author} {\bibinfo {author} {\bibfnamefont {M.~C.}\ \bibnamefont
  {Gonz\'alez-Garc\'ia}}\ and\ \bibinfo {author} {\bibfnamefont
  {M.}~\bibnamefont {Maltoni}},\ }\href {\doibase 10.1007/JHEP09(2013)152}
  {\bibfield  {journal} {\bibinfo  {journal} {JHEP}\ }\textbf {\bibinfo
  {volume} {09}},\ \bibinfo {pages} {152} (\bibinfo {year} {2013})},\ \Eprint
  {http://arxiv.org/abs/1307.3092} {arXiv:1307.3092 [hep-ph]} \BibitemShut
  {NoStop}%
\bibitem [{\citenamefont {Baryakhtar}\ \emph {et~al.}(2017)\citenamefont
  {Baryakhtar}, \citenamefont {Lasenby},\ and\ \citenamefont
  {Teo}}]{Baryakhtar:2017ngi}%
  \BibitemOpen
  \bibfield  {author} {\bibinfo {author} {\bibfnamefont {M.}~\bibnamefont
  {Baryakhtar}}, \bibinfo {author} {\bibfnamefont {R.}~\bibnamefont {Lasenby}},
  \ and\ \bibinfo {author} {\bibfnamefont {M.}~\bibnamefont {Teo}},\ }\href
  {\doibase 10.1103/PhysRevD.96.035019} {\bibfield  {journal} {\bibinfo
  {journal} {Phys. Rev. D}\ }\textbf {\bibinfo {volume} {96}},\ \bibinfo
  {pages} {035019} (\bibinfo {year} {2017})},\ \Eprint
  {http://arxiv.org/abs/1704.05081} {arXiv:1704.05081 [hep-ph]} \BibitemShut
  {NoStop}%
\bibitem [{\citenamefont {Arkani-Hamed}\ \emph {et~al.}(2007)\citenamefont
  {Arkani-Hamed}, \citenamefont {Motl}, \citenamefont {Nicolis},\ and\
  \citenamefont {Vafa}}]{Arkani-Hamed:2006emk}%
  \BibitemOpen
  \bibfield  {author} {\bibinfo {author} {\bibfnamefont {N.}~\bibnamefont
  {Arkani-Hamed}}, \bibinfo {author} {\bibfnamefont {L.}~\bibnamefont {Motl}},
  \bibinfo {author} {\bibfnamefont {A.}~\bibnamefont {Nicolis}}, \ and\
  \bibinfo {author} {\bibfnamefont {C.}~\bibnamefont {Vafa}},\ }\href {\doibase
  10.1088/1126-6708/2007/06/060} {\bibfield  {journal} {\bibinfo  {journal}
  {JHEP}\ }\textbf {\bibinfo {volume} {06}},\ \bibinfo {pages} {060} (\bibinfo
  {year} {2007})},\ \Eprint {http://arxiv.org/abs/hep-th/0601001}
  {arXiv:hep-th/0601001} \BibitemShut {NoStop}%
\end{thebibliography}%

\end{document}